\documentclass[prd,showpacs,amsmath,amssymb,nobibnotes,nofootinbib,11pt]{revtex4}
\usepackage{latexsym,epsfig}
\usepackage{bm}

\DeclareMathOperator{\sech}{sech}

\begin{document}

\title{Self-dual solutions of Yang-Mills theory on Euclidean AdS space} 

\author{{\"O}zg{\"u}r Sar{\i}o\u{g}lu}
\email{sarioglu@metu.edu.tr}
\affiliation{Department of Physics, Faculty of Arts and  Sciences,\\
             Middle East Technical University, 06531, Ankara, Turkey}

\author{Bayram Tekin}
\email{btekin@metu.edu.tr}
\affiliation{Department of Physics, Faculty of Arts and  Sciences,\\
             Middle East Technical University, 06531, Ankara, Turkey}

\date{\today}

\begin{abstract}
We find non-trivial, time-dependent solutions of the (anti) self-dual 
Yang-Mills equations in the four dimensional Euclidean Anti-de Sitter space. 
In contrast to the Euclidean flat space, the action depends on the moduli 
parameters and the charge can take any non-integer value.
\end{abstract}


\maketitle

\section{\label{intro} Introduction}

Finite action self-dual solutions with integer topological charge 
(instantons) of the Euclidean Yang-Mills (YM) theory in flat space 
($\mathbb{R}^4$), and their tunneling interpretation between the classical 
minima (in fact zeros) of the potential is well established. [See \cite{shifman} 
which compiles the original articles.] Once we depart from the Euclidean flat 
space, self-dual solutions are often drastically modified, if they are not totally
wiped out. For example, on a four dimensional hypertorus $T^4$ \cite{thooft}, 
one has many different possibilities with non-integer Pontryagin number 
(topological charge) depending on the boundary conditions on the gauge fields.
[For instantons on Taub-NUT space, see \cite{cherkis} and on 
$\mathbb{H}^{3} \times \mathbb{R}$, see \cite{harland}.]

There is, of course, a good motivation to depart from flat space and 
study self-dual YM theory in various curved backgrounds. For example, 
to define the finite temperature theory, one works on $S^1 \times \mathbb{R}^3$ 
\cite{har}. The resulting self-dual solutions are called (untwisted) calorons, which 
come with integer topological charges and provide a tunneling interpretation 
\cite{dunnetekin} in the Weyl gauge. [It is not clear if the ``twisted" 
calorons of \cite{lee,vanbaal} allow for such an interpretation.] On a generic 
four dimensional Riemannian manifold, one can obtain some general statements 
\cite{atiyah}, but one needs the explicit form of the solutions to actually utilize 
the self-dual solutions in physical problems beyond the semi-classical region.

Obviously the most relevant curved spaces are the ones that appear as solutions to 
General Relativity, with or without a cosmological constant. In principle, the effect
of gravity on the perturbative sector of quantum field theories is expected to be 
quite weak, but this need not be so in the non-perturbative sector. Gravity usually 
brings in length scales and may also introduce new topologies other than that of the 
flat space, which in turn affects the non-perturbative solutions. This said, 
on a quantum mechanical system (not a field theoretical one), one does not expect 
gravity to have much effect on tunneling. For example, one can consider the case of 
the one dimensional double well potential $V(x) = (x^2-1)^2$ as a toy model of tunneling. 
Adding a constant gravitational potential $V_{g}(z) = -mgz$, turns it into a two 
dimensional tunneling problem. However the change is not dramatic: There will be new
paths for tunneling. On the other hand, when one is interested in the vacuum of a field 
theory, such as YM theory, the effect of gravity becomes highly non-trivial. Arguably, 
the most relevant example is the YM theory in the Euclidean Schwarzschild background. 
It was shown in \cite{tekin} that all the previously obtained solutions \cite{charap1,charap2} 
are static (i.e. there is no dependence on the Euclidean time) which give rise to a 
constant potential. Thus they are solitons (monopoles and dyons) and are not instantons. 
[See \cite{radu} for more recent work.] It was conjectured in \cite{tekin} that there are 
no YM instantons with a time dependent potential in the Euclidean Schwarzschild background. 
[See \cite{etesi} for related work.]

In this paper, we shall present self-dual solutions to the $SU(2)$ YM theory 
in the four dimensional Euclidean Anti-de Sitter space\footnote{See \cite{ferstl} 
for the three dimensional version of this problem.}. [Euclidean de Sitter space can 
also be treated in the way we do here, however with two major modifications: In this 
space, time is compactified and one should stick to the region of the space inside the 
cosmological horizon.] In earlier works \cite{bout}, {\it time-independent} solutions 
were constructed, but here we will present time-dependent solutions that do have a 
non-constant YM potential.

YM theory in four dimensions is conformal, thus the self-duality equations are intact 
under a conformal scaling of the metric. Hence a naive approach would yield that in AdS 
(which is conformal to the flat space) the usual instanton solutions are pretty much 
intact and no serious modifications are to be expected. However, this is not correct since 
the AdS space is in fact conformal to the unit ball, which means that the boundary is 
at a finite distance for timelike geodesics (of course, we really have ``Euclidean" 
time here, but it is clear that the boundary effects will be quite important). Before 
we explicitly study how the boundary effects modify the topological charge of the solution, 
let us note that it has been known for a longtime that the finite action self-dual solutions 
are not necessarily classified by integer topological charge: In \cite{palla}, it was shown that 
fractionally charged (specifically charge-3/2) instantons exist if one removes the condition 
on the continuity of the group-valued function $g$, for which YM connection on $\mathbb{R}^4$ 
asymptotically becomes $A \to g^{-1}dg$. Besides the continuity assumption, one also 
assumes that \( g(\vec{x}) \to 1 \) as \(|\vec{x}| \to \infty\) leading to a compactification 
of $\mathbb{R}^3$ to $S^3$, and immediately making it transparent (for the usual instantons) that 
one has integer topological charge corresponding to the winding number of the maps 
\( g: S^3 \to S^3 \). As argued in \cite{farhi}, such a boundary condition is quite 
natural for the flat Euclidean space, but it need not be so in other spaces. AdS is one 
such example where the existence of the boundary at a finite distance completely modifies 
the instanton solutions \cite{verbin,malda}.

The outline of the paper is as follows: In section \ref{sdesss}, we set the stage for static, 
spherically symmetric Euclidean spaces and derive the self-duality equations for the 
$SU(2)$ YM theory in this background. Section \ref{swagc} is devoted to obtaining the formal
solutions of the self-duality equations for the general case, whereas subsections \ref{weyl}
and \ref{lorenz} deal with the form that these solutions take when the Weyl and the Lorenz
gauge conditions, respectively, are employed. The expressions for the topological charge and 
the potential are given in section \ref{charpot}. In section \ref{sols} we start by 
discussing how the vacuum is constructed in the Euclidean AdS space. We construct meron-like
solutions and examine their physical properties in subsection \ref{mls}. We next study the
continuous charge solutions numerically and explain their general features in subsection
\ref{ccsol}. Finally we conclude with section \ref{concs}.

\section{\label{sdesss} Self-duality on Euclidean spherically symmetric spaces}

We consider static, spherically symmetric Euclidean spaces in Schwarzschild coordinates:
\[ ds^2 = H(r) dt^2 + \frac{dr^2}{H(r)} + r^2 \, (d\theta^2 + \sin^2{\theta} \, d\phi^2) . \] 
We take the $SU(2)$ Yang-Mills theory with the standard spherically symmetric instanton 
ansatz for the gauge connection \cite{witten}
\begin{equation}
A = \frac{\tau^{a}}{2} \Big( \frac{x_{a}}{r} A_{0} \, dt + \frac{x_{a} x_{i}}{r^2} A_{1} \, dx^{i}
+ \frac{\phi_{1}}{r} \Big( \delta_{ai} - \frac{x_{a} x_{i}}{r^2} \Big) dx^{i} 
+ \epsilon_{aij} \frac{\phi_{2}-1}{r^{2}} x^{i} dx^{j} \Big) \, , \label{aanzats}
\end{equation}
where $\tau^{a}$ are the Pauli matrices. The four functions $A_{0}, A_{1}, \phi_{1}$ and 
$\phi_{2}$ depend on $t,r$ only. It is important to note that a choice of gauge at this stage 
(such as \(x^j A_j^a = A_1 = 0 \)) is not very convenient since this might lead to ostensibly 
time-dependent solutions, even though they yield a constant YM potential \cite{tekin}.

The four dimensional YM action can be reduced to a two dimensional Abelian Higgs model
in a curved background as 
\begin{equation}
I = \int_{M} d^{4} x \; \mbox{tr} \, (F \wedge *F) = 4 \pi \int_{\Sigma} d^{2} x \, \sqrt{\gamma} \,
\Big( \gamma^{\mu\nu} D_{\mu} \phi_{a} \, D_{\nu} \phi_{a} 
+ \frac{1}{4} \gamma^{\mu\alpha} \gamma^{\nu\beta} F_{\mu\nu} F_{\alpha\beta} 
+ \frac{1}{2} (1 - \phi_{a}^{2} )^{2} \Big) \, , \label{act}
\end{equation}
where spacetime indices $\mu, \nu$ refer to $(t,r)$ and $a, b$ run over $1, 2$; 
\( F_{\mu\nu} = \partial_{\mu} A_{\nu} - \partial_{\nu} A_{\mu} \) and 
\( D_{\mu} \phi_{a} = \partial_{\mu} \phi_{a} + \epsilon_{ab} A_{\mu} \phi_{b} \)
denote the two dimensional Abelian field strength and the covariant derivative, respectively. 
Here $\Sigma$ stands for some suitable region, depending on $H(r)$, in the upper half plane 
with the metric:
\[ ds^2 = \gamma_{\mu\nu} dx^{\mu} dx^{\nu} = \frac{H(r)}{r^2} dt^2 + \frac{dr^2}{r^2 H(r)} \, . \]
This type of reduction is of course well-known. [See \cite{jaffe} and the references therein.] 
One can work with the Abelian Higgs model without any loss of generality as guaranteed by 
Palais' symmetric criticality \cite{palais}. Either directly from the Abelian Higgs model or from 
the original four dimensional YM theory, the (anti) self-duality equations $F = \epsilon *F$ lead to
\begin{eqnarray}
\dot{A_{1}} - A_{0}^{\prime} & = & - \epsilon \frac{1}{r^2} (1 - \phi_{1}^{2} - \phi_{2}^{2}) \, , \label{eq1} \\
\dot{\phi_{2}} - A_{0} \, \phi_{1} & = & \epsilon H(r) (\phi_{1}^{\prime} + A_{1} \, \phi_{2}) \, , \label{eq2} \\
\dot{\phi_{1}} + A_{0} \, \phi_{2} & = & -\epsilon H(r) (\phi_{2}^{\prime} - A_{1} \, \phi_{1}) \, , \label{eq3}
\end{eqnarray}
where $\epsilon = +1$ yields the self-dual and $\epsilon = -1$ the anti self-dual choice. Here we 
have denoted derivatives with respect to $t$ and $r$ with an overdot and a prime, respectively.

Before we move on the solutions of these equations, let us note that the left-over $U(1)$ symmetry
of this model comes from the $SU(2)$ gauge transformations of the specific form 
\( U(\hat{x},t) = \exp{\big( -i (\lambda(r,t)/2) \, \hat{x} \cdot \vec{\tau} \big)} \). The effect
of this gauge transformation on $A_{\mu}$ and $\phi_{a}$ is clear. As a side remark, let us also
note that once this left-over symmetry is employed in eliminating one of these four functions,
the remaining three, which depend on $(t,r)$, define a surface in a three dimensional space. A close
scrutiny shows that for the flat space case of $H(r)=1$, the (anti) self-dual equations
(\ref{eq1})--(\ref{eq3}) are equivalent to the equations describing minimal surfaces in every
aspect, including the topological charge \cite{comtet,bay1}. For $H(r) \neq 1$, the corresponding
surfaces are not minimal in the $(t,r)$ coordinates \cite{bout}.

\section{\label{swagc} The solution without a gauge choice}

One can treat (\ref{eq2}) and (\ref{eq3}) as a linear system of equations for $A_{0}$ and
$A_{1}$ to quickly solve for these as
\[ A_{1} = \epsilon \frac{\dot{f}}{H(r) f} - \frac{g^{\prime}}{1+g^{2}} \quad \mbox{and} \quad
   A_{0} = - \frac{\dot{g}}{1+g^{2}} - \epsilon H(r) \frac{f^{\prime}}{f} \, , \]
by defining $f^{2} \equiv \phi_{1}^{2} + \phi_{2}^{2}$ and $g \equiv \phi_{1}/\phi_{2}$. 
Using these in (\ref{eq1}) then yields
\begin{equation}
 \ddot{\omega} + H(r) (H(r) \omega^{\prime})^{\prime} = \frac{H(r)}{r^2} (e^{2 \omega} - 1) \, , \label{omeq}
\end{equation}
where we have also defined $\omega \equiv \ln{f} = \ln{\sqrt{\phi_{1}^{2} + \phi_{2}^{2}}}$. The
form of (\ref{omeq}) hints at the Liouville equation which also shows up in the flat space
choice $H(r)=1$ \cite{witten}. In what follows, we will solve (\ref{omeq}) by making some
redefinitions and introducing new variables.

Now let $\omega = N(t,r) + h(r)$, where $h(r)$ is to be chosen. Then (\ref{omeq}) becomes
\begin{equation}
 \ddot{N} + H(r) (H(r) N^{\prime})^{\prime} = \frac{H(r)}{r^2} e^{2(N+h)} 
- H(r) \Big( \frac{1}{r^2} + (H(r) h^{\prime})^{\prime} \Big) \, . \label{Neq}
\end{equation}
So given $H(r)$, one can choose $h(r)$ such that \( H(r) h^{\prime} = c + 1/r \), for 
some integration constant $c$, to get rid of the last term in (\ref{Neq}). Moreover, if one 
introduces a new variable $\rho = \rho(r)$ such that \( d\rho/dr = 1/H(r) \), then 
\( \partial N/\partial \rho = H(r) N^{\prime} \) and 
\( \partial^{2} N/\partial \rho^{2} = H(r) (H(r) N^{\prime})^{\prime} \) in general.
Thus, employing such a $\rho(r)$, the left hand side of (\ref{Neq}) becomes
\( \partial^{2} N/\partial t^{2} + \partial^{2} N/\partial \rho^{2} \, . \)

Having Euclidean (A)dS space in mind, let us now introduce $\kappa = \pm 1$ 
(independent of $\epsilon$) and take \( H(r)=1 - \kappa r^2/\ell^2 \). Following the 
steps outlined above, one then finds
\[ h(r) = \left\{ \begin{array}{ll}
c \ell \tanh^{-1}{(r/\ell)} + \ln{(r/\sqrt{r^2 - \ell^2})} + k \, , & \kappa = +1 \\
c \ell \tan^{-1}{(r/\ell)} + \ln{(r/\sqrt{r^2 + \ell^2})} + k \, , & \kappa = -1 
\end{array} \, , \right. 
\quad \mbox{with a new integration constant} \; k \,. \]
We can choose the constants $c$ and $k$ by keeping in mind that as $\ell \to \infty$,
$h(r) \to \ln{r}$ to recover the flat space result. This forces us to set $c=0$
and $k=\ln{\ell}$. [Obviously this argument is valid for the AdS case. One has to
work with purely imaginary $k$ in the dS case.] Moreover, one now has
\begin{equation} 
\rho(r) =  \left\{ \begin{array}{ll}
\ell \tanh^{-1}{(r/\ell)} \, , & \kappa = +1 \\
\ell \tan^{-1}{(r/\ell)} \, , & \kappa = -1
\end{array} \, , \right. \label{rhodef}
\end{equation}
transforming (\ref{Neq}) to the celebrated Liouville equation
\begin{equation} 
\frac{\partial^{2} N}{\partial t^{2}} + \frac{\partial^{2} N}{\partial \rho^{2}} =
- \frac{\kappa}{\ell^2} e^{2 N(t,\rho)} \, , \label{Liou}
\end{equation}
whose most general solution is 
\begin{equation}
N(t,\rho) = \ln{ \Big( \frac{2 \, \ell \, |d \Phi(z)/dz|}{1 + \kappa |\Phi(z)|^{2}} \Big) } \, , \label{Lsol}
\end{equation}
where $\Phi(z)$ is an arbitrary analytic function of its complex argument $z=\rho(r)+it$ such
that \( d \Phi(z)/dz \neq 0 \). Using these, one thus obtains
\begin{equation} 
f^{2}(t,r) = \phi_{1}^{2} + \phi_{2}^{2} = 
\frac{4 \, \ell^2 \, r^2 \, |d \Phi(z)/dz|^{2}}{(r^2 - \kappa \ell^2)(1 + \kappa |\Phi(z)|^{2})^{2}} \, . 
\label{fkare}
\end{equation}

\subsection{\label{weyl} The Weyl Gauge}

So far, we have not used the gauge invariance of the action (or the field equations). 
[See the paragraph containing equation (11) of \cite{tekin} for details.] Fixing the gauge, 
we can find the unknown functions. For example, employing the Weyl gauge, $A_{0}=0$, one 
gets the following equation for the unknown $g$:
\[ \frac{\partial (\tan^{-1}{(g)})}{\partial t} = 
- \epsilon H(r) \frac{\partial (\ln{f(t,r)})}{\partial r} \, , \]
which can be solved explicitly given $\Phi(z)$. Using the solution $g(t,r)$, $A_{1}$ is
found as 
\[ A_{1} = \frac{\epsilon}{H(r)} \frac{\partial (\ln{f(t,r)})}{\partial t}
 - \frac{\partial (\tan^{-1}{(g)})}{\partial r} \, . \]
Likewise, one obtains
\[ \phi_{1} = \frac{f g}{\sqrt{1+g^{2}}} \quad \mbox{and} \quad \phi_{2} = \frac{f}{\sqrt{1+g^{2}}} \, . \]

In the Hamiltonian processes, such as tunneling, the Weyl gauge is quite useful. However,
in what follows, we will mainly work in the Lorenz gauge 
\( \partial_{\mu} ( \sqrt{\gamma} A^{\mu} ) = 0 \),
which is somewhat more convenient in finding the solutions.

\subsection{\label{lorenz} The Lorenz Gauge}

From now on we will employ the Lorenz gauge and concentrate only on $\kappa=-1$, i.e. the 
case of AdS space. Using $\sqrt{\gamma} =1/r^{2}$, the Lorenz gauge condition can be solved 
easily as \( A^{0} = -\epsilon r^{2} \chi^{\prime} \) and \( A^{1} = \epsilon r^{2} \dot{\chi} \) 
for some function $\chi(t,r)$. Defining $\psi_{a}$ as \( \phi_{a} = e^{\chi} \psi_{a} \) now 
reduces the system (\ref{eq1})--(\ref{eq3}) to
\begin{eqnarray}
\frac{\ddot{\chi}}{H(r)} + (H(r) \chi^{\prime})^{\prime} & = & 
 \frac{1}{r^2} \Big( e^{2\chi} (\psi_{1}^{2} + \psi_{2}^{2}) - 1 \Big) \,, \label{new1} \\
\dot{\psi_{2}} & = & \epsilon H(r) \psi_{1}^{\prime} \,, \label{new2} \\
\dot{\psi_{1}} & = & - \epsilon H(r) \psi_{2}^{\prime} \,, \label{new3}
\end{eqnarray}
respectively. If one further introduces a new variable $\rho=\rho(r)$ as before, such that
\( d\rho/dr = 1/H(r) \), and $\psi(z) = \psi_{1} + i \psi_{2}$, where $z = \rho + i \epsilon t$,
then (\ref{new2}) and (\ref{new3}) can be thought of as the Cauchy-Riemann conditions
that $\psi(z)$ has to satisfy to be analytic. Now when \( H(r)=1 + r^2/\ell^2 \), $\rho(r)$
is given by (\ref{rhodef}). The remaining equation (\ref{new1}) becomes
\[ \frac{\partial^{2} \chi}{\partial t^{2}} + \frac{\partial^{2} \chi}{\partial \rho^{2}} =
(\Omega(\rho))^{2} \, (e^{2 \chi(t,\rho)} \, |\psi|^{2} - 1) \,, \quad \mbox{where} \quad
1/\Omega(\rho) \equiv \ell \sin{(\rho/\ell)} \, . \]
Note that \( (\partial_{t}^{2} + \partial_{\rho}^{2}) \ln{|\psi|^{2}} = 0 \) for any
analytic function $\psi$ of $z = \rho + i \epsilon t$, except at isolated singularities.
Moreover, \( \partial_{\rho}^{2} \ln{(\ell \Omega(\rho))} = (\Omega(\rho))^{2} \). 
Using this freedom, we can set 
\[ \chi(t,\rho) = - \frac{1}{2} \ln{|\psi|^{2}} - \ln{(\ell \Omega(\rho))} + N(t,\rho) \]
to arrive at the Liouville equation (\ref{Liou}) (recall that we have $\kappa=-1$) with 
the generic solution (\ref{Lsol}). Finally one has
\[ \chi(t,\rho) = \ln{ \Big( \frac{2 \, |d \Phi(z)/dz|}
{(1 - |\Phi(z)|^{2}) \, \Omega(\rho) \, |\psi|} \Big) } \, . \]
Now the question is how to choose the analytic function $\psi(z)$. Guided by the flat space
analysis of \cite{witten}, we set $\psi(z) = d \Phi(z)/dz$ for now. One then finds
\begin{equation} 
\phi_{1}^{2} + \phi_{2}^{2} = \frac{4 \, |d \Phi(z)/dz|^{2}}
{(\Omega(\rho))^{2} (1 - |\Phi(z)|^{2})^{2}} \,, \label{ara}
\end{equation}
which is consistent with (\ref{fkare}).

\section{\label{charpot} The charge and the potential}

Before we move on the construction of the explicit solutions, let us write down the charge
and the potential in terms of the reduced fields, taking into account the boundary terms. These
will be necessary for the discussion of the physical properties of the solutions.

Defining $\epsilon_{tr} = 1/r^2$ and $\epsilon_{12} = 1$, the field equations coming from the 
variation of the 2-dimensional action (\ref{act}) are
\begin{eqnarray}
D_{\mu} \phi_{a} & = & - \gamma_{\mu\alpha} \, \epsilon^{\alpha\nu} \, \epsilon_{ab} D_{\nu} \phi_{b} \, , \\
F_{\mu\nu} & = & - \epsilon_{\mu\nu} (1 - \phi_{a}^{2}) \, ,
\end{eqnarray}
which are identical to the set (\ref{eq1})--(\ref{eq3}) for the choice $\epsilon = +1$. Using 
these in the action (\ref{act}) and taking the boundary term
\[ - 4 \pi \int_{\Sigma} d^{2} x \partial_{\alpha} \, (\sqrt{\gamma} \epsilon^{\nu\alpha} 
\epsilon_{ab} \phi_{b} D_{\nu} \phi_{a}) \to 0 \]
at $r \to \infty$, one finds
\[ I = 4 \pi \int_{\Sigma} d^{2} x \sqrt{\gamma} (1 - \phi_{a}^{2}) \, . \]
The topological charge is thus given as
\begin{equation}
Q = \frac{1}{8 \pi^{2}} \int_{M} d^{4} x \; \mbox{tr} \, (F \wedge F) = \frac{I}{8 \pi^{2}} =
 \frac{1}{2 \pi} \int_{-\infty}^{\infty} dt \, \int_{0}^{\infty} dr \, 
\frac{1}{r^2} (1 - \phi_{1}^{2} - \phi_{2}^{2}) \, . \label{cha}
\end{equation}
To really appreciate the physics of the solutions obtained, one also needs the gauge invariant 
YM potential which reads \cite{dunnetekin}
\begin{equation}
 V(t) = 2 \pi \int_{0}^{\infty} dr \, \Big( 2 H(r) (\phi_{1}^{\prime} + A_{1} \, \phi_{2})^{2} 
 + 2 H(r) (\phi_{2}^{\prime} - A_{1} \, \phi_{1})^{2}
 + \frac{1}{r^2} (1 - \phi_{1}^{2} - \phi_{2}^{2})^{2} \Big) \, . \label{pot}
\end{equation}

\section{\label{sols} The solutions}

We have seen in section \ref{lorenz} that, given an analytic function $\Phi(z)$, one can construct 
a gauge field $A$ (\ref{aanzats}) which is (anti) self-dual. However, not all (anti) self-dual 
solutions will have finite action. For example, following \cite{witten}, consider the 
meromorphic function that leads to the vacuum in flat space 
\[ \Phi(z) = \frac{a-z}{\bar{a}+z} \, , \quad \mbox{where} 
\quad a \in \mathbb{C} \quad \mbox{and for which} \quad 
\frac{d\Phi}{dz} = - \frac{2 \, \Re{(a)}}{(\bar{a}+z)^{2}} \, . \]
This choice of $\Phi(z)$ also gives a self-dual solution in Euclidean AdS space. In fact,
$\Phi(z)$ above yields
\[ f^{2}(t,r) = \frac{r^2}{(r^2+\ell^2) (\tan^{-1}(r/\ell))^{2}} \]
and 
\[ Q = \int_{-\infty}^{\infty} dt \int_{0}^{\infty} dr \, \frac{1}{r^2} (1 - f^{2}) = 
\int_{-\infty}^{\infty} dt \, \frac{2}{\ell \pi} \, , \]
which is clearly divergent. Hence, not all $\Phi(z)$ is allowed. We have to consider only
those analytic functions that lead to finite action (or charge) solutions. 

In search of these analytic functions, the representation of the vacuum plays a crucial
role. In flat space, once the vacuum is properly represented, multi-instanton solutions
can be obtained simply by taking the suitable products of the $\Phi(z)$ that corresponds
to it, i.e. they are obtained from \( \prod_{i=1}^{k} \frac{a_{i}-z}{\bar{a}_{i}+z}. \)
Just as in the flat space case, the vacuum in our setting is clearly given by 
$\phi_{1}^{2} + \phi_{2}^{2} = 1$, for which (\ref{cha}) yields $Q=0$ for the charge. 
Clearly $\phi_{2}=1$ and $\phi_{1}=A_{0}=A_{1}=0$ is the trivial vacuum $A=0$ (\ref{aanzats}). 
However, finding the analytic function $\Phi_{v}(z)$ (where the subscript `$v$' refers to 
the vacuum) that gives $A=0$ is somewhat non-trivial. From (\ref{ara}), we find that the 
function $\Phi_{v}(z)$ can be chosen as
\begin{equation} 
\Phi_{v}(z) = - \tan{\big( \frac{z}{2 \ell} - \frac{\pi}{4} \big)} =
\frac{1 - \tan{(z/2 \ell)}}{1 + \tan{(z/2 \ell)}} \,, \label{vac}
\end{equation}
which is analytic everywhere except at $z = 2 \ell \pi (k + 3/4)$ for 
$k \in \mathbb{Z}$. Moreover, one has 
\[ \lim_{\ell \to \infty} \Phi_{v}(z) = 1 \,, \quad \mbox{and} \quad
|\Phi_{v}|^{2} = \frac{\sin^{2}(-\pi/4+\rho/2 \ell) + \sinh^{2}(t/2 \ell)}
{\cos^{2}(-\pi/4+\rho/2 \ell) + \sinh^{2}(t/2 \ell)} \,. \]

In fact, one can dress up this function with a complex parameter $a$, thanks to
the invariance of the solutions (\ref{ara}) under the M{\"o}bius transformation
\[ \Phi(z) \to \frac{c + \Phi(z)}{\bar{c} \Phi(z) + 1} \quad \mbox{for any} \;\; c \in \mathbb{C} \,,\]
to get
\begin{equation} 
\tilde{\Phi}_{v}(z) = \frac{a - \tan{(z/2 \ell)}}{\bar{a} + \tan{(z/2 \ell)}} 
\quad \mbox{with} \quad a \in \mathbb{C} \,. \label{gvac}
\end{equation}
The latter yields
\[ \phi_{1} + i \phi_{2} = e^{\chi} \, \frac{d \tilde{\Phi}_{v}}{dz} = - \frac{\bar{F}}{F} \,, \]
where \( F(z) = \bar{a} \cos{(z/2 \ell)} + \sin{(z/2 \ell)} \). Let us now show that
$\tilde{\Phi}_{v}(z)$ is the function that leads to the trivial vacuum $A=0$. Recalling that 
even within the Lorenz gauge, one still has the freedom of choosing $\psi(z)$ in the solution for
$\chi(t,\rho)$, we set $\psi(z) = w(z) \, d \tilde{\Phi}_{v}(z)/dz$, where $w(z)$ 
is an analytic function of $z$, leading to 
\[ \phi_{1} + i \phi_{2} \to (\phi_{1} + i \phi_{2}) \frac{w(z)}{|w(z)|} \, . \]
Setting \( w(z) = -i F^{2} \) clearly gets one to the vacuum 
\( \phi_{1} + i \phi_{2} = i \). Thus (\ref{gvac}) gives the vacuum in the Euclidean AdS space.

As in the flat space, we will construct the finite action solutions using (\ref{gvac}). 
However, in contrast to flat space, here the action depends on the parameters
$a_{i}$ in a non-trivial way. This is to be expected since we are in AdS space
with an intrinsic length scale. [Recall that in flat space, the parameters $a_{i}$
determine the size and the locations of the instantons \cite{witten}. In AdS, the
existence of the boundary at a finite distance (as explained in the penultimate paragraph
of section \ref{intro}) drastically modifies the dependence of the action on the instanton 
moduli.] Hence the following function
\[ \prod_{i=1}^{k} \frac{a_{i}-\tan{(z/2 \ell)}}{\bar{a}_{i}+\tan{(z/2 \ell)}} \]
leads to a finite action self-dual solution. The topological charge and the action depend 
on the $a_{i}$ in a non-trivial way and, unfortunately, the action can only be
calculated numerically for generic $a_{i}$.

\subsection{\label{mls} The meron-like solutions}
In the special case of $a_{i}=1$, the calculations can be carried out analytically. For
example, consider $\Phi(z) = (\Phi_{v}(z))^{2}$. Then it is not hard to show that (\ref{ara}) yields
\[ \phi_{1}^{2} + \phi_{2}^{2} = \frac{4 |\Phi_{v}|^{2}}{(1+|\Phi_{v}|^{2})^{2}}, \quad \mbox{and} 
\quad 1 - \phi_{1}^{2} - \phi_{2}^{2} = \Big( \frac{\cos{(\rho/\ell-\pi/2)}}{\cosh{(t/\ell)}} \Big)^{2} \, .\]
Using this in (\ref{cha}) gives $Q=1/2$ since
\[ \int_{-\infty}^{\infty} dt \, \int_{0}^{\infty} dr \, 
\frac{1}{r^{2}} (1 - \phi_{1}^{2} - \phi_{2}^{2}) = \pi \, . \]
Similarly, one can also work out the potential in this case. From (\ref{pot}) it simply reads
\[ V(t) = \frac{3 \pi^{2}}{2 \ell} \sech^{4}{(t/\ell)} \,. \]

Other examples of half-integer charges can be constructed in this vein, however 
calculations get rather complicated. We were able to show that if one chooses
$\Phi(z) = (\Phi_{v}(z))^{n}$, where $n=2^{k}$ for $k \in \mathbb{Z}^{+}$, then $Q=(n-1)/2$.
Note that these are genuinely new and non-trivial solutions in the Euclidean
AdS space and completely disappear in the flat space limit $\ell \to \infty$.

In flat space, charge-1/2 solutions of the \emph{full} YM equations exist and go under the
name as `merons' \cite{fubini}. Note however that these are singular solutions with a divergent
action. Additionally, note also that charge-3/2 \emph{self-dual} solutions in flat space were 
constructed as well \cite{palla}. Here, we have shown that the Euclidean AdS space admits similar 
half-integer meron-like solutions with a finite action.

\subsection{\label{ccsol} The continuous charge solutions}

Let us now consider more general solutions. Let
\[ \Phi(z) = (\tilde{\Phi}_{v}(z))^{2} 
= \Big( \frac{a-\tan{(z/2 \ell)}}{\bar{a}+\tan{(z/2 \ell)}} \Big)^{2} \]
with $a \equiv \alpha + i \beta$, a complex parameter. Then the relevant integrand
for the action, charge or the potential energy reads
\begin{equation} 
\frac{1}{r^{2}} (1 - \phi_{1}^{2} - \phi_{2}^{2}) =
 \frac{16 \alpha^{2} \sin^{2}{(\rho/\ell)} \, |\cos{(z/2 \ell)}|^{4}}
{r^{2} \Big( 4 \alpha^{2} \, |\cos{(z/2 \ell)}|^{4} + \sin^{2}{(\rho/\ell)} 
+ \big( \sinh{(t/\ell)} - 2 \beta \, |\cos{(z/2 \ell)}|^{2} \big)^{2} \Big)^{2}} , \label{integ}
\end{equation}
where \( |\cos{(z/2 \ell)}|^{2} = \cosh^{2}{(t/2\ell)} - \sin^{2}{(\rho/2 \ell)} \). Unfortunately,
the computations from this point on can only be performed numerically. In figures 1 and 2, we
have calculated the topological charge $Q$ and the potential $V(t)$, respectively, 
to exhibit the general features of the solutions obtained using (\ref{integ}).

Fig. 1 depicts the topological charge $Q$ as a function of $\alpha$ and $\beta$. When
$\alpha=0$, $Q=0$ as expected since $\alpha$ (as discussed below) is a parameter that gives
the scale of the solution, and when $\alpha=0$, the solution becomes trivial. Even though it
is not very apparent in Fig. 1, the charge $Q$ changes very slowly with $\beta$. Note that
as long as $\ell \neq 0$ or $\infty$, $Q$ does not depend on $\ell$. [Note that the discontinuity
at $\alpha=0$ is natural, since $\alpha \to 0$ limit is formally equivalent to the $\ell \to \infty$
limit for which we have $Q \to 1$, which is the flat space result. However, when $\alpha$ is
exactly zero to start with, a careful analysis gives $Q=0$ as explained above.]
\begin{figure}[ht]
\begin{center}
\epsfig{file=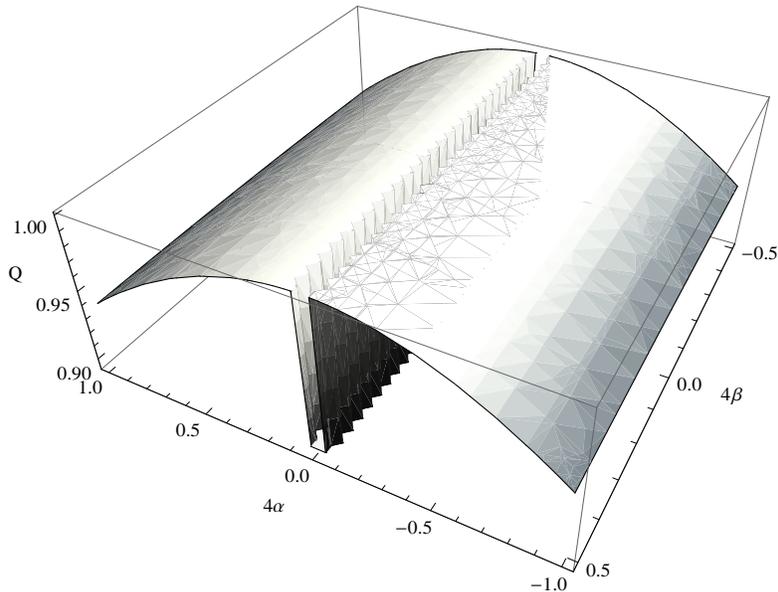,height=7.8cm,width=10.4cm} 
\caption{The topological charge $Q$ when $\ell=2$.}
\end{center}
\end{figure}

In Fig. 2, we have plotted the potential $V(t)$ as a function of time $t$ for fixed $\ell=2$. As
argued below, $\beta$ determines the `location' of the solution on the time $t$-axis.
\begin{figure}[ht]
\begin{center}
\epsfig{file=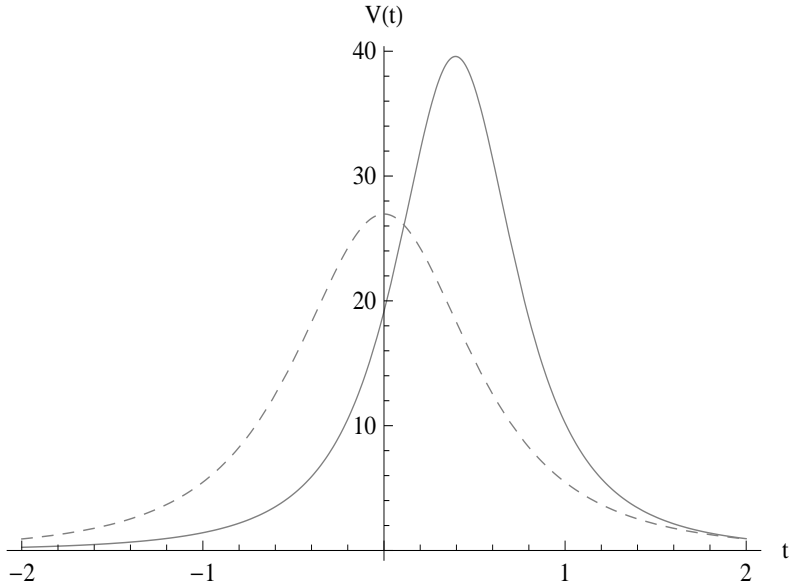,height=7.8cm,width=10.4cm} 
\caption{The potential $V(t)$ as a function of $t$ when $\ell=2$. The solid line is for $\alpha=1/8$, 
$\beta=1/10$; whereas the dashed line is for $\alpha=3/16$ and $\beta=0$.}
\end{center}
\end{figure}

It is worth emphasizing that in the flat space limit $\ell \to \infty$, one obtains 
the corresponding flat space solution after redefining $a \to 2 \ell a$. [That is why we
have $4 \alpha$ and $4 \beta$ in the labels of the axes in Fig. 1.] The interpretation of the 
parameters $a_i$, in terms of the `scale' and the location (on the $t$-axis), follows the
discussion in flat space. Because of our choice, $a_i$ are dimensionless of course, but 
$\ell$ clearly acts as the proper length parameter. Let us define the `location' of the solution 
as the point on the time $t$-axis where the potential energy takes its maximum value . [For 
multi-instantons, the maxima of the potential define the individual locations of the `instantons'.] 
From (\ref{integ}), it follows that for such maxima one should look for the solutions of 
\( \tilde{\Phi}_{v}(z) \frac{d \tilde{\Phi}_{v}(z)}{dz} = 0 .\) One can check that this boils down to 
finding the zeros of \( \tilde{\Phi}_{v}(z) = 0 \), since its derivative does not vanish in the 
relevant domain. Hence, one should solve \( \alpha + i \beta = \tan{( (\rho_0 + i t_0)/2 \ell)} \), 
which yields
\begin{equation}
\alpha = \frac{\tan{(\rho_0/2 \ell)} \, \sech^2{(t_0/2 \ell)}}
{1 + \tan^2{(\rho_0/2 \ell)} \, \tanh^2{(t_0/2 \ell)}} \,, \qquad \qquad 
\beta = \frac{\tanh{(t_0/2 \ell)} \, \sec^{2}{(\rho_0/2 \ell)}}
{1 + \tan^2{(\rho_0/2 \ell)} \, \tanh^2{(t_0/2 \ell)}} \,. \label{sizloc}
\end{equation}
Given $\alpha$ and $\beta$, one immediately finds the scale $\rho_0$ and the location $t_0$. 
It is important to note that unlike the flat space case, here $\alpha$ and $\beta$ are restricted 
to the domains $\alpha \in [0,1]$ and $\beta \in [-1,1]$. This follows from (\ref{sizloc}) by
a careful consideration of the ranges of $\rho \in [0, \pi \ell/2)$ and $t \in (-\infty,\infty)$.
Specifically, consider $\Phi(z) = (\Phi_{v}(z))^{2}$, i.e. $\alpha = 1$ and $\beta = 0$
case, studied in subsection \ref{mls}. This corresponds to the case of having $t_{0}=0$ and
the size of the `meron' going to infinity, that is the `meron' fills the whole space.

To keep the discussion simple, we have refrained from considering either
\[ \Phi(z) = \prod_{i=1}^{k} \frac{a_{i}-\tan{(z/2 \ell)}}{\bar{a}_{i}+\tan{(z/2 \ell)}} \,, \]
or higher powers of $\tilde{\Phi}_{v}(z)$, i.e. \( \Phi(z) = (\tilde{\Phi}_{v}(z))^{k} \)
with $k \geq 3$, here. Except for the $a=1$ case, which is studied in subsection \ref{mls}, 
we have not been able to compute either the charge or the potential analytically.
However, it is clear by construction that these also lead to (anti) self-dual solutions
and in principle it is possible to numerically obtain the physical properties of these as well.

\section{\label{concs} Conclusions}

We have studied the (anti) self-dual $SU(2)$ gauge fields in the Euclidean
Anti-de Sitter space. We have shown that the problem eventually reduces to 
finding the solutions of the Liouville equation on the strip 
\( 0 \leq \Re{(z)} < \pi \ell/2 \) of the complex plane. We have seen that 
given \emph{any} analytic function, one can construct (anti) self-dual solutions 
which do not in general have finite action. Finding finite action (or charge) solutions 
reduces to finding \emph{proper} analytic functions inside the relevant strip
as discussed in section \ref{sols}. The solutions we have found have quite 
interesting properties: They can have any non-integer charge including fractional 
values. Our solutions depend on the time coordinate $t$ and have non-trivial 
YM potential. In this respect, they are quite distinct than the earlier, 
\emph{static} solutions \cite{bout}. Looking at the potential, one can see that 
the solutions presented here resemble pretty much the flat space instantons, 
having \( V(t \to \pm \infty) = 0 \) and a bump (or bumps in between). We have also 
explained how a non-integer charge is quite natural in the AdS context. [Note that 
even in flat space, non-integer charge values are allowed \cite{farhi,palla}.]

In the search of the solutions, we have left one question unanswered: Are there generic 
integer-charge solutions? There seems to be no compelling reason why there should not 
be any. Unfortunately though, we have not been able to find these solutions. It is 
quite interesting that certain fractionally charged solutions appear more naturally 
in AdS than the integer ones. A further direction of research would be to consider 
the Euclidean de Sitter (dS) space. It is clear that most of the equations in this paper 
also work for the dS space. The problem arises again in finding the proper analytic 
functions that will yield finite action solutions. In dS, because of the cosmological 
horizon, one has to search for time-periodic solutions, namely finite temperature caloron 
solutions, restricted to live inside the horizon.

\section{\label{ackno} Acknowledgments}
This work is partially supported by the Scientific and Technological Research
Council of Turkey (T{\"U}B\.{I}TAK). B.T. is also partially supported by the 
T{\"U}B\.{I}TAK Kariyer Grant 104T177.


\begin{thebibliography}{99}

\bibitem{shifman}
  M.A. Shifman,
  ``Instantons in gauge theories,''
  %\href{http://www.slac.stanford.edu/spires/find/hep/www?irn=3917819}
  {\it Singapore, Singapore: World Scientific (1994) 488 p}.

\bibitem{thooft}
  G.'t Hooft,
  %``Some Twisted Selfdual Solutions For The Yang-Mills Equations On A Hypertorus,''
  Commun. Math. Phys. {\bf 81}, 267 (1981).

\bibitem{cherkis}
  S.A. Cherkis,
  ``Instantons on the Taub-NUT Space,''
  arXiv:0902.4724 [hep-th].

\bibitem{harland}
  D. Harland,
  %``Hyperbolic calorons, monopoles, and instantons,''
  Commun.\ Math.\ Phys.\  {\bf 280}, 727 (2008)
  [arXiv:hep-th/0703277].

\bibitem{har}
  B.J. Harrington and H.K. Shepard,
  %``Periodic Euclidean Solutions And The Finite Temperature Yang-Mills Gas,''
  Phys. Rev. D {\bf 17}, 2122 (1978).

\bibitem{dunnetekin}
  G.V. Dunne and B. Tekin,
  %``Calorons in Weyl gauge,''
  Phys. Rev. D {\bf 63}, 085004 (2001) 
  [arXiv:hep-th/0011169].

\bibitem{lee}
  K. Lee and C. Lu,
  %``SU(2) calorons and magnetic monopoles,''
  Phys. Rev. D {\bf 58}, 025011 (1998)
  [arXiv:hep-th/9802108].

\bibitem{vanbaal}
  T.C. Kraan and P. van Baal,
  %``Periodic instantons with non-trivial holonomy,''
  Nucl. Phys. B {\bf 533}, 627 (1998)
  [arXiv:hep-th/9805168].

\bibitem{atiyah}
  M.F. Atiyah, N.J. Hitchin and I.M. Singer,
  %``Selfduality In Four-Dimensional Riemannian Geometry,''
  Proc. Roy. Soc. Lond. A {\bf 362}, 425 (1978).

\bibitem{tekin}
  B. Tekin,
  %``Yang-Mills solutions on Euclidean Schwarzschild space,''
  Phys. Rev. D {\bf 65}, 084035 (2002) 
  [arXiv:hep-th/0201050].

\bibitem{charap1}
  J.M. Charap and M.J. Duff,
  %``Gravitational Effects On Yang-Mills Topology,''
  Phys. Lett. B {\bf 69}, 445 (1977).

\bibitem{charap2}
  J.M. Charap and M.J. Duff,
  %``Space-Time Topology And A New Class Of Yang-Mills Instanton,''
  Phys. Lett. B {\bf 71}, 219 (1977).

\bibitem{radu} 
  E. Radu, D.H. Tchrakian and Y. Yang,
  %``Spherically symmetric selfdual Yang-Mills instantons on curved 
  %backgrounds in all even dimensions,''
  Phys. Rev. D {\bf 77}, 044017 (2008)
  [arXiv:0707.1270 [hep-th]].

\bibitem{etesi}
  G. Etesi and M. Jardim,
  %``Moduli spaces of self-dual connections over asymptotically locally flat gravitational instantons,''
  Commun. Math. Phys. {\bf 280}, 285 (2008)
  [arXiv:math/0608597].

\bibitem{ferstl} 
  A. Ferstl, B. Tekin and V. Weir,
  %``Gravitating instantons in 3 dimensional anti de Sitter space,''
  Phys. Rev. D {\bf 62}, 064003 (2000)
  [arXiv:hep-th/0002019].

\bibitem{bout}
  H. Boutaleb-Joutei, A. Chakrabarti and A.Comtet,
  %``Gauge Field Configurations In Curved Space-Times,''
  Phys. Rev. D {\bf 20}, 1884 (1979).

\bibitem{palla}
  P. Forgacs, Z. Horvath and L. Palla,
  %``An Exact Fractionally Charged Selfdual Solution,''
  Phys. Rev. Lett. {\bf 46}, 392 (1981).

\bibitem{farhi}
  E. Farhi, V.V. Khoze and R. Singleton,
  %``Minkowski space non-Abelian classical solutions with noninteger winding number change,''
  Phys. Rev. D {\bf 47}, 5551 (1993)
  [arXiv:hep-ph/9212239].

\bibitem{verbin}
  Y. Verbin,
  %``GRAVITATIONAL SUPPRESSION OF VACUUM TUNNELING IN YANG-MILLS THEORIES,''
  Class. Quant. Grav. {\bf 7}, L89 (1990).

\bibitem{malda}
  J.M. Maldacena and L. Maoz,
  %``Wormholes in AdS,''
  JHEP {\bf 0402}, 053 (2004)
  [arXiv:hep-th/0401024].

\bibitem{witten}
  E. Witten,
  %``Some exact multipseudoparticle solutions of classical Yang-Mills theory,''
  Phys. Rev. Lett. {\bf 38}, 121 (1977).

\bibitem{jaffe} 
  A.M. Jaffe and C.H. Taubes,
  ``Vortices And Monopoles. Structure Of Static Gauge Theories,''
  %\href{http://www.slac.stanford.edu/spires/find/hep/www?irn=826421}{SPIRES entry}
  {\it Boston, USA: Birkhaeuser (1980) 287 p. (Progress In Physics, 2)}.

\bibitem{palais}
  R.S. Palais,
  Commun. Math. Phys. {\bf 69}, 19 (1979).

\bibitem{comtet}
  A. Comtet,
  %``Instantons And Minimal Surfaces,''
  Phys. Rev. D {\bf 18}, 3890 (1978).

\bibitem{bay1}
  B. Tekin,
  %``Multi-instantons in R**4 and minimal surfaces in R**(2,1),''
  JHEP {\bf 0008}, 049 (2000) 
  [arXiv:hep-th/0006135].

\bibitem{fubini}
  V.de Alfaro, S. Fubini and G. Furlan,
  %``A New Classical Solution Of The Yang-Mills Field Equations,''
  Phys. Lett. B {\bf 65}, 163 (1976).

\end{thebibliography}
\end{document}